\RequirePackage{fix-cm}
\documentclass[twocolumn]{svjour3}          
\smartqed  
\usepackage{graphicx}
\usepackage{subcaption}
\usepackage{caption}
\usepackage{threeparttable}
\usepackage{booktabs}
\usepackage[dvipsnames]{xcolor}
\RequirePackage{hyperref}

\def\makeheadbox{\relax}

\begin{document}

\title{Cryptocurrencies Activity as a Complex Network: Analysis of Transactions Graphs\footnotemark}

\author{Luca Serena \and
        Stefano Ferretti \and 
        Gabriele D'Angelo
}


\institute{S. Ferretti \at
              University of Urbino (Italy) \\
              \email{stefano.ferretti@uniurb.it}
}

\maketitle

\footnotetext{The publisher version of this paper is available at \url{https://doi.org/10.1007/s12083-021-01220-4}.
\textbf{{\color{red} This is the authors' version of the following
article: ``Luca Serena, Stefano Ferretti, Gabriele D'Angelo. Cryptocurrencies Activity as a Complex Network: Analysis of Transactions Graphs. Peer to Peer Networking and Applications (Springer)''.}}}

\begin{abstract}
The number of users approaching the world of cryptocurrencies exploded in the last years, and consequently the daily interactions on their underlying distributed ledgers have intensified. In this paper, we analyze the flow of these digital transactions in a certain period of time, trying to discover important insights on the typical use of these technologies by studying, through complex network theory, the patterns of interactions in four prominent and different Distributed Ledger Technologies (DLTs), namely Bitcoin, DogeCoin, Ethereum, Ripple. In particular, we describe the Distributed Ledger Network Analyzer (DiLeNA), a software tool for the investigation of the transactions network recorded in DLTs. We show that studying the network characteristics and peculiarities is of paramount importance, in order to understand how users interact in the DLT. For instance, our analyses reveal that all transaction graphs exhibit small world properties.
\end{abstract}


\maketitle

\section{Introduction}
Cryptocurrencies have been a disruptive innovation in the world of economic transactions. Based on cryptography and on the decentralization of the information, these virtual tokens are thought to offer an alternative to the traditional fiat currencies. Unlike fiat currencies, however, cryptocurrencies base their value exclusively on the trust of the investors: except those implementations that base their value on traditional financial assets (i.e. ~stablecoins like Tether or Security Token Offerings), no central banks and monetary reserves can influence the supply of money and, as such, indirectly control inflation. Also for these motivations, often the economical value of the cryptocurrencies is subject to high fluctuations.

Bitcoin, launched in 2009, was the ancestor of all the other cryptocurrencies and it is even now by far the most popular and the most used one~\cite{nakamoto2019bitcoin}. In view of the success of Bitcoin, many other cryptocurrencies were created in the last decade: according to \cite{coinmarketcap}, currently there are over $3500$ types of cryptocurrencies, with the global cryptomarket worth almost $400$ billion dollars. However, as reported by \cite{coinmarketcap}, Bitcoin still holds 64\% of the cryptomarket capitalization, with Ethereum steadily in the second place with the 11\%.

The cryptomarket is currently still growing and there are different reasons why one might be fascinated by this technology: some users might be interested in the anonymity (or pseudo-anonymity) features~\cite{michael2018blockchain}, others in the lack of central entities in charge of managing the money transfers, in the economic value of the cryptocurrencies built over these ledgers (for example, for speculation purposes), or finally, with the aim to build decentralized applications that are able to exploit the features offered by smart contracts~\cite{luu2016making}.

Blockchains and cryptocurrencies have been widely studied in these last years. Mostly, the domains of investigation are related to security issues~\cite{smart-sec,DAHIYA2021193}, usage in Internet of Things deployments and drones~\cite{IJCAC.2019040104,9354926}, Smart Cities environments~\cite{ESPOSITO2021102468,TAN2021103517,LI2019432}, large-scale heterogeneous networks~\cite{SHI2020}, as well as on their economical impact~\cite{li2017technology}.

In this paper, we follow a totally different approach, in fact we perform a study of these technologies by relying on mechanisms that are typical of the (social) networks analysis (e.g.~\cite{ugander2011anatomy,gabielkov2014studying}). Thus, we study the interactions among the accounts involved in the transactions that are recorded on the Distributed Ledger Technologies (DLTs). For this purpose, the system to be studied is treated as a graph, where links are drawn when pairs of accounts have a some sort of interaction. This kind of analysis is viable, since unlike normal methods for exchanging money, the whole set of transactions is traced in the distributed ledger and furthermore it is visible to everyone~\cite{sf-gda}.

To this aim, we devised and implemented a novel software tool called DiLeNa (Distributed Ledger Network Analyzer)\footnote{An early version of this work appeared in~\cite{10.1145/3410699.3411361}. This paper is an extensively revised and extended version where more than 50\% is new material.}. 
This software is able to grab the transactions stored in the distributed ledger of different DLTs, create an abstraction of a network and then measure some important related metrics, which allow to extract some peculiar characteristics of the network.
We claim that these metrics can provide insights on the respective technologies and on the use that users make of them.
Specifically, in our work we apply this approach to four different DLTs, namely Bitcoin, DogeCoin, Ethereum and Ripple. The analysis on their transactions graphs reveals that the networks -some more, some less- feature a small world behaviour, and that most of the nodes have very few connections, even though some hubs with a very high degree exist. The exhibited behaviour is supposed to be influenced by the considered time interval and by the implementation features of the DLTs, so different types of transactions graphs have been taken into account. We think that the approach we propose in this paper, should help to get more information about how interactions through DLTs occur and that it could have useful applications, such as for anti-money laundering purposes.

The remainder of the paper is organized as follows. Section 2 introduces some background and related work. Section 3 describes the design choices of the software tool and deals with the critical aspects of its implementation. Section 4 analyzes the results obtained by analyzing the some relevant distributed ledgers. Finally, Sections 5 provides some concluding remarks.

\section{Background and Related Work}\label{sec:back}
In this section, some background that is essential to understand the rest of the paper is introduced. Specifically, the topics covered will be the DLT technology, the representation of complex systems as graphs and some metrics that can be used in order to evaluate the specific characteristics of a network (e.g.~small world property).

\subsection{DLT and Blockchain Technologies}
Cryptocurrencies, unlike traditional banking systems, store their data into distributed ledgers, thus making the information decentralized and avoiding single points of failure. DLTs work as distributed and immutable databases and the nodes involved in the management and in the update of the DLTs check the integrity and the consistency of the data, ensuring the correctness of the system. Most of the DLTs are permissionless, meaning that no prior approval to actively participate in the system is needed.

There can be different implementations of a distributed ledger but most of the cryptocurrencies rely on the blockchain, a data structure that stores the transactions into containers called blocks, logically linked among each other through the use of cryptographic techniques. All the cryptocurrencies require a consensus strategy that allows all the nodes of the system to agree about the actual state of the distributed ledger. Also in this case, different schemes are available, but the most popular implementations use the so-called Proof of Work (PoW), requiring to solve computationally intensive crypto-puzzles in order to validate the blocks and the transactions contained therein. The act of solving the cryptographic puzzles is called ``mining''. Bitcoin was the first cryptocurrency launched in the market and still now is the most famous and the most used one.  Recently, also Ethereum gained popularity because it allows to execute, other than simple transactions, actual contracts written with code, the so called ``smart contracts''.

In most of the systems, users are identified with addresses derived from their public cryptographic key and there is no trivial way to associate the addresses with the real identity of the users. So it is possible (and often happens) that certain users control multiple accounts and that multiple addresses are linked to a single account.

\subsection{Graphs and Complex Systems}

A graph can be an effective way to represent systems or parts of systems where different entities interact. Specifically, a graph $G$ is a data structure defined as $G=(V,E)$ where $V$ is a set of entities called nodes or vertices and $E$ is a set of edges, which are links representing connections between pairs of nodes. In directed graphs, the link relation is not necessarily commutative, while in undirected graphs the edges symbolize links in both directions: that means that if $A$ is connected with $B$ then $B$ is also connected with $A$. Graphs can also be either weighted or unweighted, depending if the edges are marked with weights, i.e.~numerical values that are able to represent different measures depending on the meaning of the graph such as distances, economical costs, number of interactions, etc.

In a graph, it is not always possible to reach each couple of nodes by following a path along the edges. A subgraph including all the nodes that can communicate with each other, through a certain path, is called a ``connected component''. There can be various connected components in a graph, and usually the component with the largest number of nodes is referred to as the ``main component''. In directed graphs a component can either be strongly or weakly connected: in the first case all the couples of nodes can communicate in both directions, in the second case the communication might be possible just in one direction.

Depending on the problem, the meaning of a graph can be different. In a map, for example, vertices may represent some locations and the weights of the edges are the distance between two nodes~\cite{sf-gda,ferretti2017modeling,FERRETTI20131631}. In our case, the vertices are addresses of a certain blockchain and the presence of an edge indicates that there has been an interaction between the two nodes (i.e.~a transaction).

\subsection{Network Topologies}
The structure of a network can be described through some mathematical properties of the associate graph. Here, we focus on two specific graph topologies, i.e.~\emph{random graphs} and \emph{small world graphs}.

The random graphs are networks completely generated by random processes and where no presence of hubs or skewed distribution is expected to occur. A typical method to generate a synthetic random graph is the Erdos-Renyi model~\cite{paul1959random}: it is an algorithm that takes as arguments (i) the number of nodes to be created and (ii) either the total number of edges or the probability that a link between two nodes exists. Then, the algorithm proceeds to consult iteratively random sources to decide where edges are going to be placed.

Small world networks~\cite{watts1998collective} are a graph topology where usually two nodes are connected by a low number of hops and where often neighbor nodes share other neighbors in common. Frequently, these graphs are characterized by the presence of cliques (i.e.~subgraphs where each couple of nodes is directly linked) and hubs, which serve as connectors among highly clustered groups of nodes. Several examples of real networks that exhibit a small world structure exist, i.e. food chains, electric power grids, neural networks, telephone call graphs and social influence networks. Detecting if a graph is a small world can be useful in various application areas. For example, in medicine it can give information about how a disease spreads within a population, while in telecommunications and computer science it is possible to exploit the knowledge on the structuring of the graph in order to optimize the dissemination and the storage of data~\cite{gda-jpdc-2017}.

Two metrics are commonly used in order to evaluate if a graph has a small world property~\cite{bassett2006small}:
\begin{itemize}
    \item \emph{Average shortest path length (ASPL)}. In unweighted graphs, the shortest path is the path that connects two nodes with the minimum number of hops. The ASPL is thus calculated averaging the shortest paths among all the couples of connected nodes. The shortest path can be computed by using the classic Dijkstra's algorithm~ \cite{dijkstra1959note}, and the procedure is iterated for $n * (n-1)$ times in the worst case. The overall time complexity of the algorithm is $O(|V|^2|E|+|V|^3 log|V|)$.
    \item \emph{Average clustering coefficient (ACC)}. The clustering coefficient of a node is the fraction that indicates the percentage of the neighbors of such a node that are in turn directly linked. It basically tells how much the friends of a node (i.e.~the node neighbors) are friends among themselves. The average clustering coefficient of the whole network is obtained by averaging the clustering coefficients of all the nodes. It produces an output ranging from $0$ to $1$: the higher the value the more clustered is the graph.
\end{itemize}

The two mentioned metrics can be used to characterize a small network as follows. Such values have to be compared with a random graph, created with the same number of nodes and edges. Then, one can state that the graph features small world properties if, compared to a random graph of the same size, the average clustering coefficient is significantly higher and the average shortest path length similar (or smaller)~\cite{ferretti2017modeling}.

In practice, all this can be measured by computing the following metrics
$$\sigma ={\frac {\frac {C}{C_{r}}}{\frac {L}{L_{r}}}}$$ 
In this last formula, $C$ is the ACC of the analyzed graph and $C_r$ is the ACC of the random graph. Similarly, $L$ is the ASPL of the analyzed graph and $L_r$ is the ASPL of the random graph. The higher $\sigma$, the more pronounced the small world behaviour is.

Other graph topologies exist, for example scale-free networks are characterized by the degree distribution following a power law. In these types of graphs there are few hubs connected with a large number of nodes, while most of the nodes are scarcely connected~\cite{d2009simulation}.

\section{The DiLeNa Tool}
\begin{figure*}[t]
  \centering
    \includegraphics[width=0.9\textwidth]{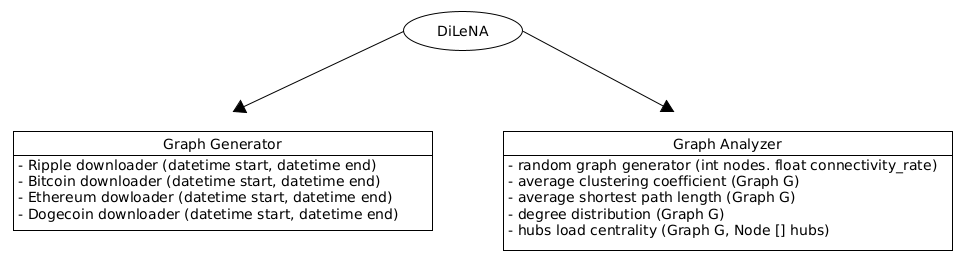}
  \caption{DiLeNA's modular design: the graph generator retrieves the transactions graph of a certain DLT and then the graph analyzer calculates the requested metrics.}
  \label{dilena}
\end{figure*}

The proposed software, that is freely available on the research group website~\cite{pads}, is modular and it is composed of two main components (as shown in Figure~\ref{dilena}):
\begin{itemize}
    \item \emph{Graph Generator}: it is in charge of downloading the transactions of the examined DLT, generated during the time interval of interest. Then, a directed graph is built, that represents the interactions among the nodes. The vertices of the graph correspond to the addresses in the DLT and, for each transaction, an edge directed from the sender to the recipient of the transactions is made (if not already existing).
    \item \emph{Graph Analyzer}: this module is in charge of calculating the typical metrics related to the obtained graph. Among the others, the tool is able to measure the degree distribution, network clustering coefficient, as well as to identify the main component and some of its main metrics, such as the average shortest path. Moreover, the tool computes if the network is a small world, by comparing it with a corresponding random graph (with the same amount of nodes and edges).
\end{itemize}

The generated graph is stored in Pajek format~\cite{pajek}. This format is particularly efficient for our scope, since it is an optimized data encoding format for graphs, that allows to save space, by mapping the addresses (represented by hexadecimal strings of about 40 characters each) with an incremental integer. For example, one day of Bitcoin transactions (e.g.~1st September 2020) needed 125.8 MB to be stored in Pajek format, while with the standard JSON representation, the amount of memory consumed was 368.6 MB.

\subsection{Graph Generator}
As already mentioned, this module is in charge of retrieving a set of transactions, by inspecting the distributed ledger related to the cryptocurrencies under study. Needless to say, every DLT stores its transactions according to a different encoding format. Moreover, each DLT typically provides its own set of specific Application Programming Interfaces (APIs) to inquiry the ledger and retrieve contents. Thus, for each DLT a different data retrieval methodology (and software module) has been implemented.

For instance, as concerns Ripple, a Python library is available, named python-ripple-lib, which allows to download up to $100$ transactions given a certain period of time. Due to such a limit, in case that larger datasets are required, multiple requests have to be performed, until the whole time interval of interest is covered. Regarding Ethereum, in a previous version of DiLeNa, the transactions were downloaded by using Infura, a service that provides access to a remote Ethereum node through APIs. However, recent restrictions on the free version of Infura lead us to lean on Etherscan.io, a block explorer and analytics platform for Ethereum. Dogecoin's transactions are retrieved from the website SoChain~\cite{sochain}, which offers the functionality of blockchain explorer. The site also shows the content of other blockchains, like ZCash, Litecoin and Dash. So extending the software for such distributed ledgers should be trivial work. Finally, in order to get Bitcoin transactions, we use the ``curl'' library to query the Blockchain.info~\cite{bexplorer} website.

More specifically, the download phase was designed to follow a parallel approach: the user can optionally define the number of concurrent workers to use ($1$ by default) in order to better share the workload for the requests of blocks and transactions.

\subsection{Graph Analyzer}
This module has been specifically implemented in Python in order to have a seamless interaction with the NetworkX library, a software package for the analysis and the manipulation of graphs and complex networks~\cite{networkx}.

Under the usage viewpoint, the Graph Analyzer enables the user to specify what graph is to be analyzed and then to compute some metrics provided by NetworkX. In particular, for our scope we are mainly interested in:
\begin{itemize}
    \item Degree Distribution;
    \item Average Clustering Coefficient (ACC);
    \item Average Shortest Path Length (ASPL) of the main component.
\end{itemize}

After completing these computations, the Graph Analyzer generates a new random graph with the same number of nodes and an equivalent number of edges using the well-known Erdos-Renyi model. This makes possible a comparison between the two graphs that is necessary to find if the networks that are analyzed have small world properties. It is worth noticing that, also in this case, we implemented the computation of the metrics enabling a parallel execution. In fact, the user has the chance to define the number of workers to be used (a single worker is used by default).

\subsection{Design and Implementation Issues}
To be able to perform the analysis described above, the first issue that needs to be properly addressed is how to automate the download of transactions from the DLTs.
Generally speaking there are two main options for downloading slices of a distributed ledger:
\begin{itemize}
    \item An API that, among other things, allows downloading blocks and/or transactions.
    \item Blocks and/or transactions are available (usually in JSON format) on certain web pages. In this case, once the indices of blocks corresponding to the desired period of time are found, all one has to do is to iterate requests to the site by using the right parameters.
\end{itemize}
There are also cryptocurrencies, however, for which none of the two methods is available. Thus, if one wanted to retrieve the data of such cryptocurrencies, it would be necessary to download the full distributed ledger, often made of hundreds of gigabytes.
Furthermore, there are some cryptocurrencies like Monero with a particular focus on privacy that prevent the observer from accessing some information. Monero is built on top of a public blockchain, but most of its portions are encrypted. Senders, recipients and amounts being transferred are hidden to third parties through the use of stealth addresses, RingCT and Bulletproofs~\cite{monero}.

Another problematic aspect concerns the parallelization of the download. In the proposed version of the software (for Ripple, Ethereum and Dogecoin) the user can specify the number of workers that will manage the retrieval of blocks and transactions. Though, it happens frequently that the server replies with a 429 HTTP status code, indicating that a rate limit mechanism is implemented and that too many requests have been issued. When this occurs, the program is paused for a few seconds before resuming its activity. Thus, it is advised to use a limited number of concurrent workers in order to not overload the servers with too many requests.

Once the transactions are downloaded, the biggest concern is how to calculate the metrics linked with the shortest path in a reasonable amount of time. For example, the average shortest path length can require up to several months to complete, while metrics such as the clustering coefficient or the degree distribution are almost immediate to compute. A parallel approach can significantly reduce the amount of time required for the metrics, sharing the workload among multiple workers (i.e.~each worker can be executed on a different CPU core). However, this may not be sufficient to get the results in an acceptable amount of time, so it could be necessary to only consider a random sample of nodes on which to calculate the shortest path length.

Another problem that arises with large-scale graphs is that their representation in a data structure could exceed the space available in RAM on the computer used for the analysis. This often requires a more complex management of the memory and a costly (in terms of time) access to the secondary storage. To mitigate this issue, the adoption of the Pajek format, to represent the generated graphs, can be really useful. In fact, it saves a lot of space with respect to the JSON format that is often used. For example, a single month of Ethereum transactions required 1.9 GBs to be stored in JSON and just 540 MBs in Pajek.

\section{Analysis of the Results}

In this section, the outcomes of our analyses are discussed. As mentioned, we will report the outcomes from  Bitcoin, DogeCoin, Ethereum and Ripple.

\subsection{Setup, Methodologies and Performance}
Our main intention was to analyze the behavior of the most popular cryptocurrencies' ledgers in a specific time-frame (i.e.~all the transactions in a specific day or month, in our case either 1st September 2020 or the entire month of September 2020). While most of the metrics do not require a lot of time to be calculated, the average shortest path length is very critical timewise. In order to mitigate this problem, we investigated the viability to consider just a sample of the nodes of the main weakly connected component, thus reducing the number of times that Dijkstra algorithm needs to be applied. For example, with a sample of $1/n$ nodes, just $1/n^2$ of the total paths of the main component need to be computed. From our experiments (Table~\ref{sample}), it turns out that the difference between the actual results and the sampled results is negligible (around 1\% of difference using a sample including 10\% of the nodes). Thus, for the sake of computation, we decided to adopt such a technique. However, even with the use of samples, the Dijkstra algorithm remains time-dependent on the number of nodes and edges of the main component, therefore calculating the average shortest path length on very big graphs always remains very time consuming.

\begin{table*}[h]
  \centering
  \caption{Analysis of the Ripple's transactions graph: one day of transactions, comparison of different samples size.}
\begin{tabular}{|c|c|c|}
\hline
\textbf{Sample Size}   &  \textbf{Main Component ASPL} & \textbf{Difference} \\
\hline
\textit{10\%}    & 4.4548 & 0.98\% \\ 
\textit{25\%}    & 4.395 &  0.37\% \\
\textit{100\%}   & 4.4116 & - \\ 
\hline
\end{tabular}
\label{sample}
\end{table*}

Different sample sizes were applied, depending on the size of the graph. For example, to study one month of Ripple transactions we used a 20\% sample. The analysis took $19$ hours to complete, 6h37m for the retrieved graph and the rest for the random graph. In this case, the generation of the random graph only lasted $10$ minutes, but in other cases it is a very long operation, requiring up to some days. The server used for conducting this analysis is equipped with an Intel Xeon CPU E3-1245 (v5 @ 3.50GHz) running Ubuntu 18.04.5 LTS.

Analyzing one day of Ethereum transactions (sample: 10\%) required $14$ hours, most of them (i.e. 11h35m) for computing the ASPL of the random graph. In general, to calculate the metrics on the random graph is much more time consuming compared to the retrieved graph. This is due to the higher ASPL, always detected in the random graphs, which directly protracts the time required to compute such a metric.

\subsection{Ripple}
Released in 2013, Ripple is both a cryptocurrency and a platform that allows, with negligible fees (only there to prevent Denial-of-Service attacks), to connect banks, payment providers and digital asset exchanges, by offering a solution for real-time money transfers that could be slow and costly due to different countries and currencies involved. Ripple was thought as a bridge currency between fiat currencies when making cross-border payments or between crypto and fiat currencies. Unlike most of the cryptocurrencies, Ripple does not use a blockchain as a distributed ledger. Transactions are stored in a network made of independent validating nodes that constantly compare their transaction records and the consensus is achieved by using the Ripple Protocol Consensus Algorithm. Ripple has the advantage of being extremely fast to validate transactions, achieving to validate up to $1500$ transactions per second~\cite{xrp-per-sec}, against the $7$ transactions per second manageable by Bitcoin in its maximum throughput~\cite{btc-scaling}. However, unlike most of the other cryptocurrencies, the system is not fully decentralized, since all the Ripple tokens are pre-mined but only some of them are available to the market, being periodically released at the discretion of the company that controls the system, Ripple Lab~\cite{xrp}. At the time of writing (i.e.~April 2021), Ripple is the forth cryptocurrency by market capitalization, only behind Bitcoin, Ethereum and Binance Coin~\cite{coinmarketcap}.

Another interesting characteristic of Ripple is that it is possible to avoid the noise caused by the presence of change addresses, created in Bitcoin-like systems when a part of the input has to be returned to the sender of a transaction. Ripple allows for different types of transactions~\cite{XRP_transactions} like the creation or the removal of accounts, payment channels and escrows. However, here we focus on payments, since they are the only type of transaction through which it is possible to directly map a sender and a receiver of economical funds.

\begin{figure}
  \centering
    \includegraphics[width=0.5\textwidth]{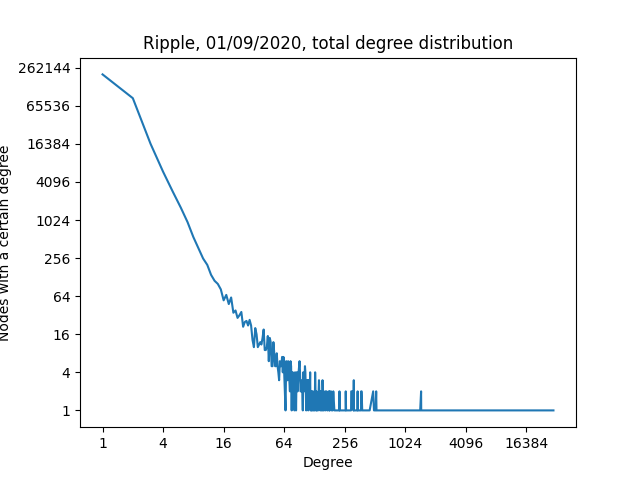}
    \includegraphics[width=0.5\textwidth]{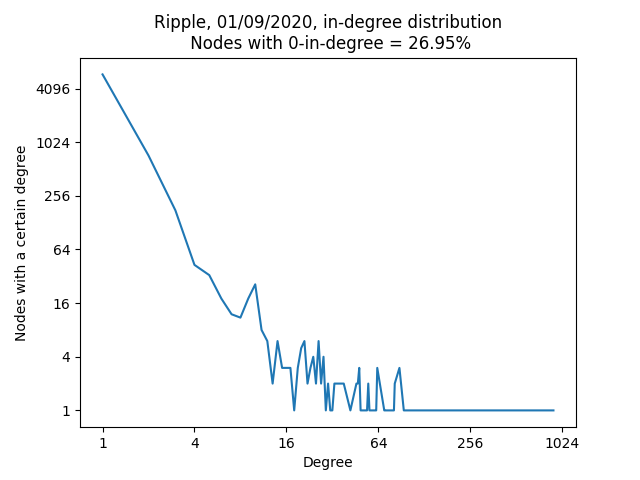}
    \includegraphics[width=0.5\textwidth]{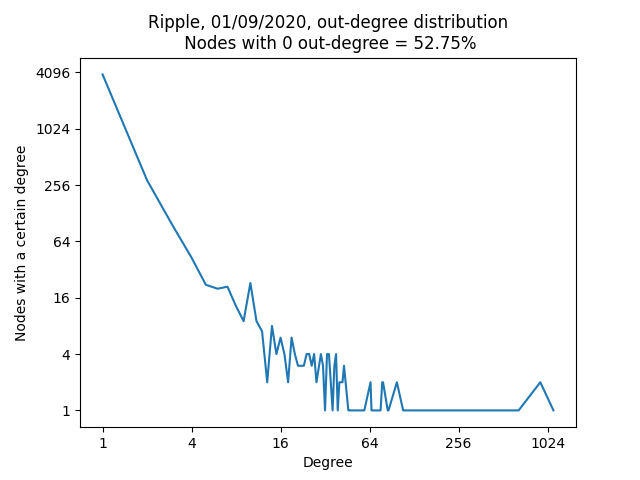}

  \caption{In, out and total degree distribution of the Ripple graph. Based on the transactions dated: 1st September 2020. The plot is performed in logarithmic scale on both axes.}
  \label{rippledegree}
\end{figure}

Figure~\ref{rippledegree} shows the degree distribution of the graph resulting from one day of Ripple transactions (September 1st, 2020). Most of the nodes interacted just sporadically, with more than $7$ out of $10$ nodes just having either incoming or outgoing edges but not both of them. It also turned out that the number of nodes with $0$ as out-degree is almost double compared to the nodes that never received a transaction. Moreover, there are few hubs with a large number of interactions. The node with highest degree (i.e.~the most connected node), for example, had connections with $1125$ nodes, that is  11.58\% of the whole network.

\begin{figure}[t]
  \centering
    \includegraphics[width=0.5\textwidth]{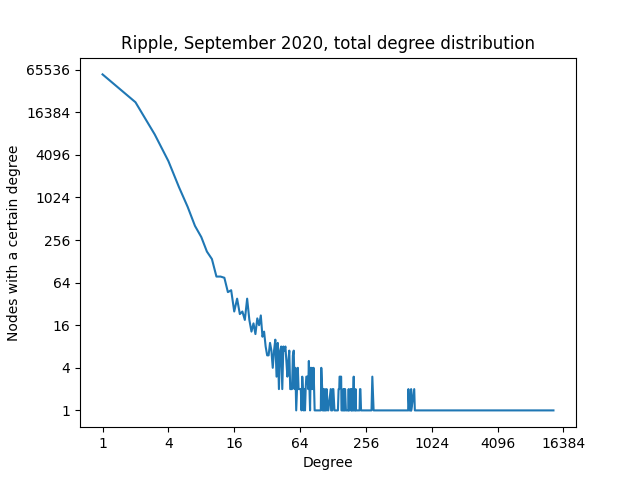}
    \includegraphics[width=0.5\textwidth]{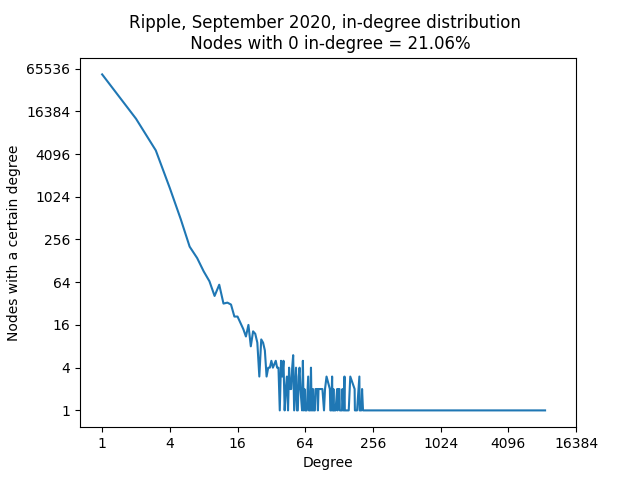}
    \includegraphics[width=0.5\textwidth]{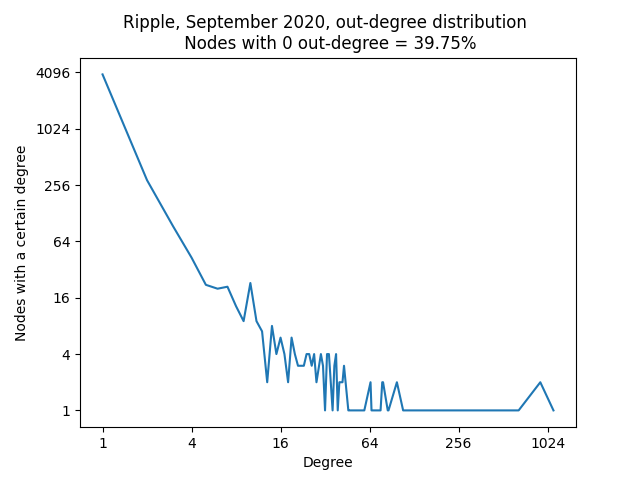}
  \caption{In, out and total degree distribution of the Ripple graph. Based on the transactions dated: September 2020. The plot is performed in logarithmic scale on both axes.}
  \label{rippledegreemonth}
\end{figure}

After this preliminary evaluation, we have collected and analyzed a full month of transactions (i.e.~September 2020). The obtained results are reported in Figure~\ref{rippledegreemonth}. It is worth noting that, as expected, there are fewer nodes with a very low degree distribution with respect to the previous graph. This is due to the fact that there are some nodes that may have a single interaction in one day, but multiple in a longer time period. As a result, the percentage of nodes with just one interaction dropped from 71.6\% (as reported in the first graph) to the 59.5\% in the second graph. Regarding the hubs, there are $4$ nodes with more than {10\,000} edges, with the most connected one having connections with 15\% of the nodes. Here 95\% of the nodes belong to the main component, a higher amount with respect to the 89.4\% reported in the one-day transactions graph. Moreover, it is interesting that the size of the one-month graph is just $9.7$ times larger (in terms of total nodes) and $6$ times larger (in terms of edges) with respect to the on-day graph. This happens because a significant part of the transactions carried out in $30$ days were performed among nodes that already directly interacted, thus no new node nor new link is added to the network. In fact, in the one-day graph, 84\% of the transactions happened between two addresses that already communicated in that direction, and this percentage rises to 93.6\% in the one-month graph.

Table~\ref{xrptable} shows the outcome of the computed metrics. We can state that the transactions graph has a small world behaviour both considering one day and one month of interactions. In the first case, the ratio of the average shortest path lengths is $0.28$ and the clustering coefficient of the Ripple graph is almost $600$ times greater compared to the random graph. In the second case the ratio of the average shortest path lengths is similar ($0.24$) and ratio of the average clustering coefficient is even considerably higher ($12512$).

The tests over one month of Ripple transactions were repeated changing configuration for the evaluation of the average shortest path. By default, the ASPL is calculated among the nodes belonging to the main weakly connected component, considering only the couple of nodes that are actually connected. A second test was made, considering the main strongly connected component, but no significant difference was noticed, except that the clustering coefficient of the main component (which includes 22.2\% of the nodes, against 95\% of the weakly connected component) is considerably higher. Finally, the test was repeated considering the graph as undirected. Here, some differences occur: first of all the clustering coefficient is higher, as a direct consequence of the increased number of links. Then, for the same reason, the ASPL is lower ($3.12$). However, taking into account the comparison with a similar random graph, also in this case we can state that the transactions graph has small world properties, even though both the ratio are slightly higher.

\begin{table*}[h]
    \caption{Analysis of the load centrality of the hubs with the highest degree in the transaction graph.}
    \begin{subtable}[h]{0.45\textwidth}
        \centering
        \begin{tabular}{|c|c|}
        \hline
        \textbf{Node Degree} & \textbf{Load Centrality} \\
        \hline
        \textit{13\,276}    & 0.1004 \\ 
        \textit{12\,352}    & 0.0447 \\ 
        \textit{11\,935}    & 0.0479 \\ 
        \textit{10\,863}    & 0.0399 \\ 
        \textit{8\,697}     & 0.0851 \\ 
        \textit{6\,860}     & 0.1147 \\ 
        \textit{4\,818}     & 0.0389 \\ 
        \textit{3\,785}     & 0.0216 \\ 
        \textit{3\,251}     & 0.0168 \\ 
        \textit{2\,753}     & 0.0194 \\ 
        \textit{2\,684}     & 0.0144 \\ 
        \hline
        \end{tabular}
        \caption{Ripple hubs' load centrality.}
        \label{load_centrality_ripple}
    \end{subtable}
    \hfill
    \begin{subtable}[h]{0.45\textwidth}
        \centering
        \begin{tabular}{|c|c|}
        \hline
        \textbf{Node Degree} & \textbf{Load         Centrality} \\
        \hline
        \textit{30\,465}    & 0 \\ 
        \textit{24\,858}    & 0.0000012 \\ 
        \textit{8\,418}    & 0.0369 \\ 
        \textit{8\,384}    & 0.000001 \\ 
        \textit{6\,992}     & 0 \\ 
        \textit{5\,514}     & 0.00607 \\ 
        \textit{5\,100}     & 0 \\ 
        \textit{5\,055}     & 0.00495 \\ 
        \textit{4\,622}     & 0.0043 \\ 
        \textit{4\,484}     & 0.000001 \\ 
        \hline
        \end{tabular}
        \caption{Ethereum hubs' load centrality.}
        \label{load_centrality_eth}
    \end{subtable}
    \begin{subtable}[h]{0.45\textwidth}
        \centering
        \begin{tabular}{|c|c|}
        \hline
        \textbf{Node Degree} & \textbf{Load         Centrality} \\
        \hline
        \textit{8\,259}    & 0.211 \\ 
        \textit{4\,601}    & 0.04047 \\ 
        \textit{2\,678}    & 0.09613 \\ 
        \textit{2\,209}    & 0.02883 \\ 
        \textit{1\,984}     & 0.02489 \\ 
        \textit{1\,765}     & 0 \\ 
        \textit{1\,143}     & 0.000008 \\ 
        \textit{797}     & 0.01693 \\ 
        \textit{716}     & 0.00588 \\ 
        \textit{608}     & 0.02464 \\ 
        \hline
        \end{tabular}
        \caption{Dogecoin hubs' load centrality.}
        \label{load_centrality_doge}
    \end{subtable}
    \hfill
    \begin{subtable}[h]{0.45\textwidth}
        \centering
        \begin{tabular}{|c|c|}
        \hline
        \textbf{Node Degree} & \textbf{Load         Centrality} \\
        \hline
        \textit{35\,597}    & 0.07518 \\ 
        \textit{4\,571}     & 0.00205 \\ 
        \textit{4\,438}     & 0.00336 \\ 
        \textit{3\,753}     & 0.000000027 \\ 
        \textit{3\,734}     & 0.00106 \\ 
        \textit{3\,573}     & 0.0031 \\ 
        \textit{3\,319}     & 0 \\ 
        \textit{3\,045}     & 0.00159 \\ 
        \textit{2\,958}     & 0.00098 \\ 
        \textit{2\,861}     & 0 \\ 
        \hline
        \end{tabular}
        \caption{Bitcoin hubs' load centrality.}
        \label{load_centrality_btc}
    \end{subtable}
\end{table*}

Finally, the load centrality was computed for the most connected hubs of the system. The load centrality of a node is the fraction of all shortest paths that pass through that node. Table~\ref{load_centrality_ripple} shows that, in general, the more the nodes are connected the more the load centrality tends to be high, even if a strict correlation cannot be claimed.

\begin{table*}[h]
  \centering
  \caption{Comparison between the Ripple graph and the equivalent random graph, one day and one month of transactions.}
\begin{tabular}{|c|c|c|c|}
\hline
\textbf{Graph}   & \textbf{Graph ACC} & \textbf{Main Comp. ASPL} & \textbf{Main Comp. ACC} \\
\hline
\textit{Ripple (one day)}    & 0.0516 & 4.4116 & 0.0561 \\ 
\textit{Random (one day)}    & 0.000089 & 16.1623 & 0.000094 \\
\textit{Ripple (one month)}    & 0.1489 & 4.4307 & 0.1562 \\
\textit{Random (one month)}    & 0.000012 & 18.5325 & 0.000012 \\
\hline
\end{tabular}
\label{xrptable}
\end{table*}

\subsection{Bitcoin}
In Bitcoin, accounts are addresses composed of alphanumeric characters, generated by applying a hash function to a generated public key, associated with a given user. The balance of the wallets is not written somewhere in the blockchain, so novel transactions must point to a set of Unspent Transaction Outputs (UTXO), in order to prove that the sender owns the necessary amount of money for the payment. Since UTXO inputs must be spent entirely, when the sum of the pointed inputs of a transaction is greater than the actual expenditure, then the unspent part of the inputs are sent back to the original owner, similarly to the change someone receives after conducting a cash transaction in a store. For change operations, often different addresses are used (the so-called change addresses)~\cite{change}, in order to enhance the privacy while making more difficult the traceability of the blockchain. In fact, there can be multiple addresses linked to a given wallet. 
Furthermore, unlike Ripple and Ethereum (where a transaction can only have a single address as sender and another one as recipient), transactions with multiple inputs and multiple outputs are allowed: in these cases, for our analysis the $n$ input addresses are mapped with the $m$ output addresses, resulting in $n * m$ links. This is possible since one wallet can have multiple receiving addresses; thus, one can collect the UTXO associated with these addresses to create a single transaction.

\begin{figure}
  \centering
    \includegraphics[width=0.5\textwidth]{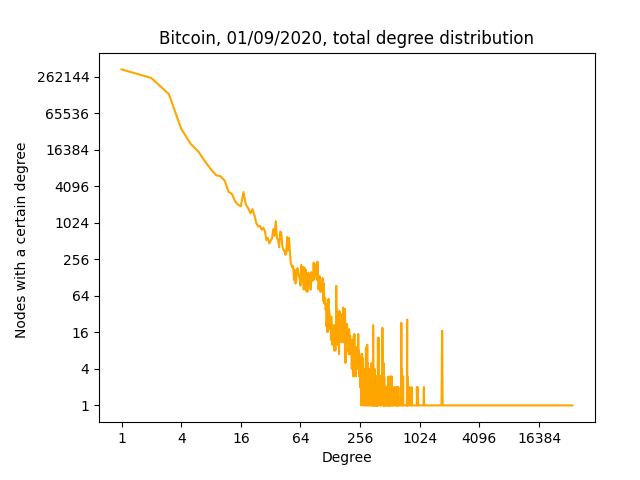}
    \includegraphics[width=0.5\textwidth]{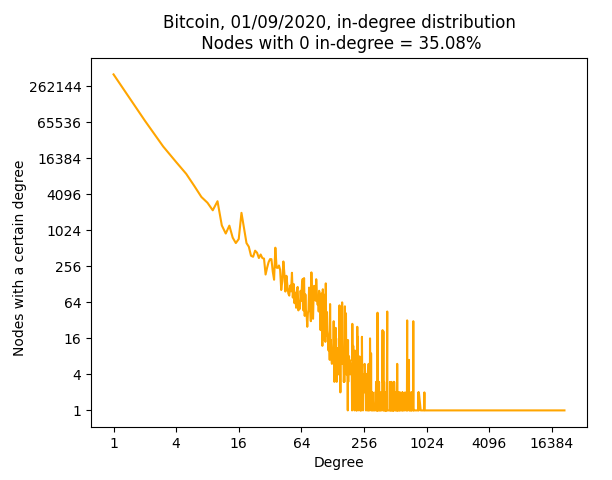}
    \includegraphics[width=0.5\textwidth]{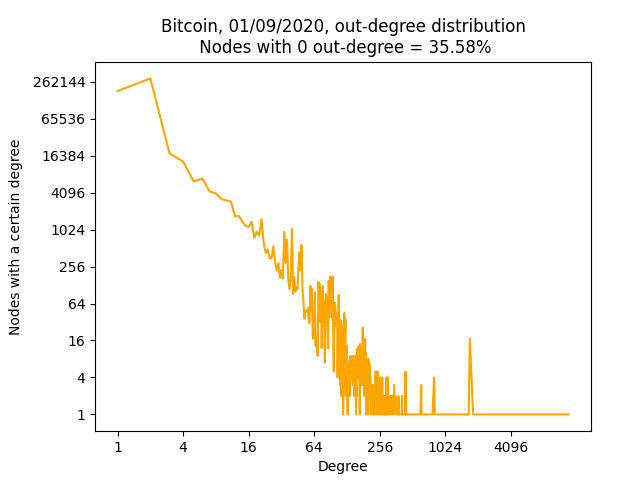}
  \caption{In, out and total degree distribution of the Bitcoin graph. Based on the transactions dated: 1st September 2020.The plot is performed in logarithmic scale on both axes.}
  \label{btcdegree}
\end{figure}

Similarly to other graphs, the analysis of the Bitcoin degree distribution shows the presence of very few hubs. However, here the percentage of active nodes involved with more than $5$ accounts in daily transactions is greater than 10\%, a significantly higher fraction with respect to the other distributed ledgers (see Figure~\ref{btcdegree}). This is probably due to the transactions with multiple addresses as input and/or output.
The peculiar thing about the Bitcoin graph is that the most connected node has a degree ($35\,597$ connections, whose $21\,729$ are incoming edges) that is almost $8$ times greater than the degree of the second most connected node. Such a node is linked with the 4\% of the other active nodes in the network.

\begin{table*}[h]
  \centering
  \caption{Comparison between the Bitcoin graph and the equivalent random graph.}
    \begin{tabular}{|c|c|c|c|}
    \hline
    \textbf{Graph}   & \textbf{Graph ACC} & \textbf{Main Component ASPL} & \textbf{Main Component ACC} \\
    \hline
    \textit{Bitcoin}    & 0.02067 & 10.0886 & 0.02358 \\ 
    \textit{Random}    & 0.00000214 & 14.7141 & 0.00000216  \\
\hline
\end{tabular}
\label{btctable}
\end{table*}

Like the previously analyzed graph, also the Bitcoin transactions graph has a small world property. However, as we can observe from Table~\ref{btctable}, the ratio of the average shortest path length of the transactions graph and the random graph is $0.69$, the highest among our analyses.

\subsection{Ethereum}
Ethereum, unlike Bitcoin, does not rely on a UTXO model to retrieve the balance of the accounts~\cite{UTXOvsAccount}. Similarly to banks, each account has a balance, which increases when it receives Ethers, and decreases when it sends Ethers to other users. Thus, there is no need for change addresses, because all the transactions simply deduct from one account and add to another. In addition, there are two types of Ethereum accounts~\cite{8345547}:
\begin{itemize}
    \item private key controlled user accounts;
    \item contract code-controlled accounts.
\end{itemize}
Each contract can be identified by its Ethereum address in the same way a normal Ethereum user can. In Ethereum, from the standpoint of the blockchain, transactions have a single address both as sender and as recipient. This differs from Bitcoin where transactions with multiple inputs and multiple outputs are allowed.

\begin{figure}
  \centering
    \includegraphics[width=0.5\textwidth]{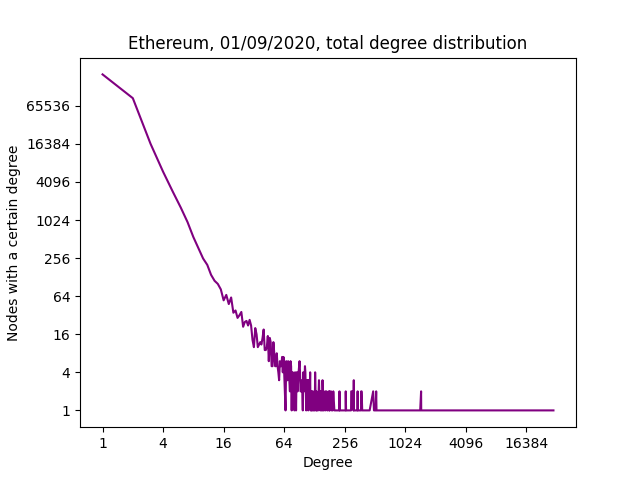}
    \includegraphics[width=0.5\textwidth]{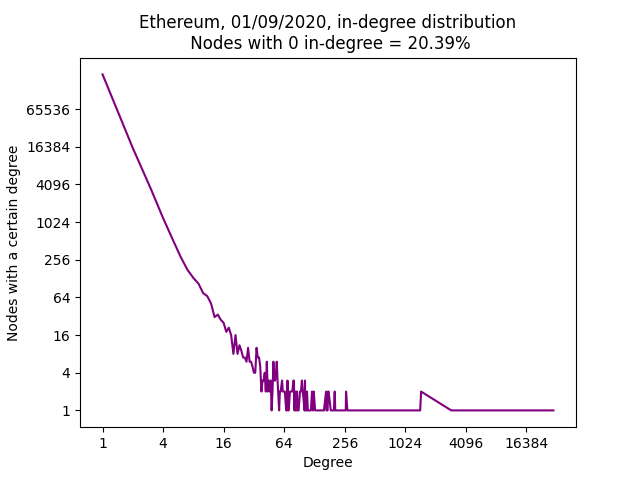}
    \includegraphics[width=0.5\textwidth]{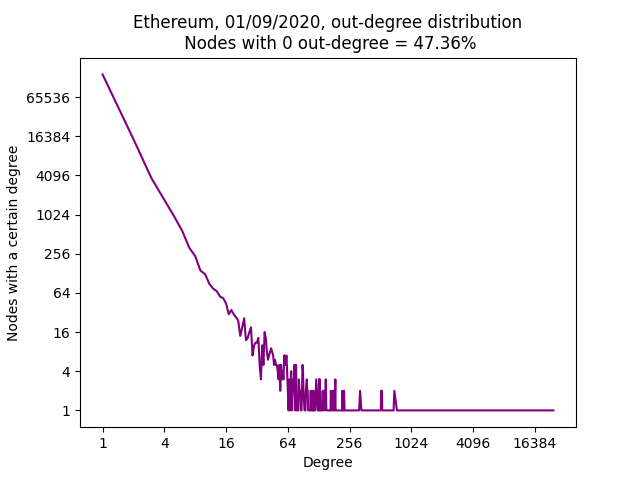}
 \caption{In, out and total degree distribution of the Ethereum graph. Based on the transactions dated: 1st September 2020.The plot is performed in logarithmic scale on both axes.}
  \label{ethdaydegree}
\end{figure}

\begin{figure}
  \centering
    \includegraphics[width=0.5\textwidth]{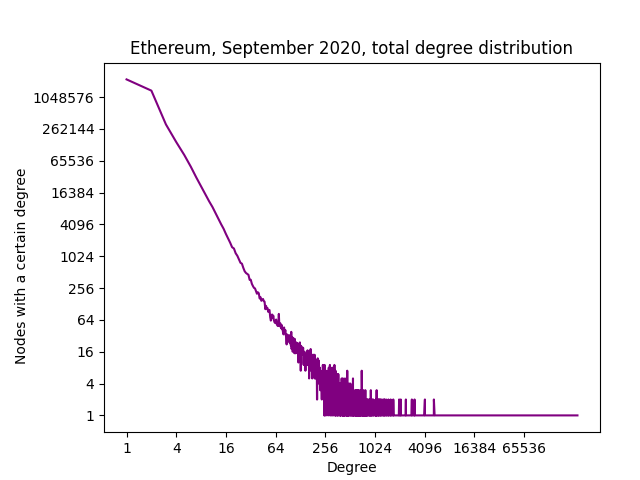}
    \includegraphics[width=0.5\textwidth]{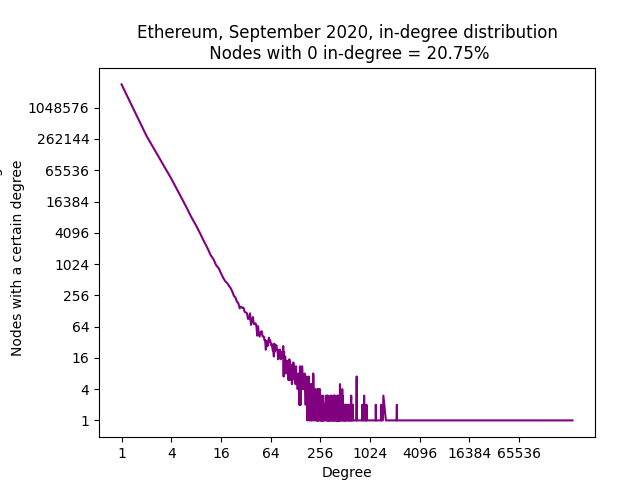}
    \includegraphics[width=0.5\textwidth]{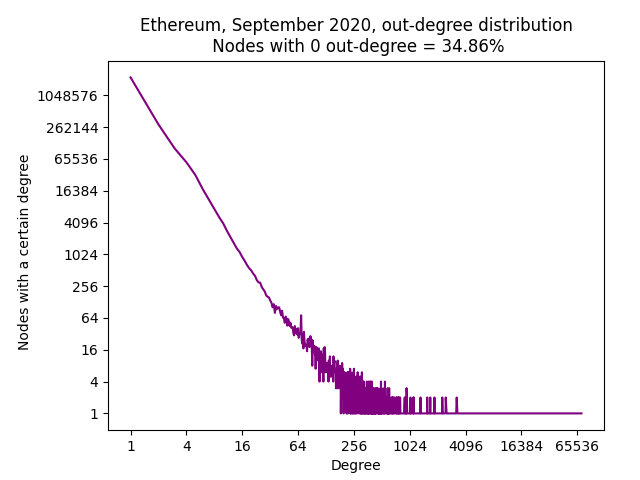}
 \caption{In, out and total degree distribution of the Ethereum graph. Based on the transactions dated: September 2020. The plot is performed in logarithmic scale on both axes.}
  \label{ethmonthdegree}
\end{figure}

Unlike Ripple, just 25.4\% of the Ethereum transactions pass through already existing edges, thus the main part of the payments are carried out among nodes that have not communicated between each other in that 24 hours interval. This percentage grows to 48.4\% taking into account the full month.
The degree distribution resulting from our analysis shows that also here there are few nodes with a very high degree, thus implying the presence of hubs. Figure~\ref{ethdaydegree} shows the degree distribution of the graph resulting from the transactions performed on 1st September 2020. With $322\,467$ nodes and just $376\,587$ edges, it turns out that 80.75\% of the nodes belongs to the main weakly connected component. Almost $2/3$ of the nodes just had interactions with another node and there are $10$ nodes with a degree greater than $4000$, including two giant hubs, the first one having incoming edges with the 9.45\% of the network and the second one having outgoing edges with 7.7\% of the nodes. Considering one month of transactions (September 2020) we have slight differences, as we can notice in Figure~\ref{ethmonthdegree}. The ratio edges-nodes is now slightly higher ($1.47$, versus $1.17$ of the one day transactions graph) and the percentage of nodes with just one interactions fell from 63.66\% to 51.59\%. The most connected hub here has interactions with just 1.16\% of the network and there are other $5$ hubs with over 1\% of the possible links.

\begin{table*}[h]
  \centering
  \caption{Comparison between the various Ethereum graphs, one day and one month of transactions.}
\begin{tabular}{|c|c|c|c|}
\hline
\textbf{Graph}   & \textbf{Graph ACC} & \textbf{Main Comp. ASPL} & \textbf{Main Comp. ACC} \\
\hline
\textit{ETH normal (one day)} & 0.00879 & 6.9129 & 0.0103 \\
\textit{ETH normal (one month)} & 0.03025 & 7.46296 & 0.03174\\
\textit{ETH internal (one day)} & 0.0.00646 & 3.9556 & 0.00686\\
\textit{ETH internal (one month)} & 0.00764 & 4.1599 & 0.00793 \\
\hline
\end{tabular}
\label{ethtable}
\end{table*}

Table~\ref{ethtable} shows that, also in this case, we can state that Ethereum transactions graph has a small world behaviour. In fact, the ratio of the average clustering coefficient between the transactions graph and the random graph is $14\,320$, and the ASPL ratio is $0.13$.
Regarding the load centrality of the hubs, unlike Ripple, very few shortest paths pass through the most connected hubs. Figure~\ref{load_centrality_eth} shows that among the ten most connected nodes, just one has a load centrality value greater than 1\%.

The website ``Etherscan.io'' allows to visualize, other than normal transactions, also the so called internal transactions, which are some kind of interactions that occurred between two smart contracts. Despite the name, they are not actually considered real transactions, since they are not directly included in the blockchain. An internal transaction is triggered whenever a smart contract needs to send Ethers, Tokens or make some sort of mechanic on its own. The degree distribution of Ethereum internal transactions resulted as particularly unbalanced: in the one-day graph 99\% of the addresses just either sent or received one transaction (99.38\% in the one-month graph) while the most clustered hub has incoming connections with 36.8\% of the nodes (41.1\% in the one-month graph). This happens despite having a similar edges-nodes ratio (around $1.1$) with respect to the Ethereum graphs with normal transactions.

\subsection{DogeCoin}
Dogecoin is a Litecoin based cryptocurrency, launched in December 2013. Similarly to Bitcoin, it features Proof-of-Work as a consensus protocol and the accounts rely on a UTXO model to get the balance of the accounts (thus several change addresses will appear). Transactions with multiple inputs and multiple outputs are allowed. 
Like Bitcoin, several input addresses in a transaction can indicate that these multiple addresses are associated to the same user.

In particular, in our analysis $55\,212$ transactions have led to $513\,759$ binary connections between addresses and to a total of $143\,641$ directed edges in the resulting graph. Most of the transactions were composed of 1-2 inputs and 1-2 outputs, however few transactions with a very big number of addresses involved were detected. For example, the largest involved $167$ input addresses and $1141$ output addresses.

Table~\ref{dogetable} shows that also Dogecoin exhibits small-world behaviour, having $840$ as clustering coefficient ratio and $0.68$ as ASPL ratio.
\begin{table*}[h]
  \centering
  \caption{Comparison between the Dogecoin graph and the equivalent random graph.}
\begin{tabular}{|c|c|c|c|}
\hline
\textbf{Graph}   & \textbf{Graph ACC} & \textbf{Main Component ASPL} & \textbf{Main Component ACC} \\
\hline
\textit{Dogecoin}    & 0.06665 & 5.1633 & 0.06819 \\ 
\textit{Random}    & 0.0000799 & 7.6305 & 0.0000763 \\
\hline
\end{tabular}
\label{dogetable}
\end{table*}

\begin{figure}
  \centering
    \includegraphics[width=0.5\textwidth]{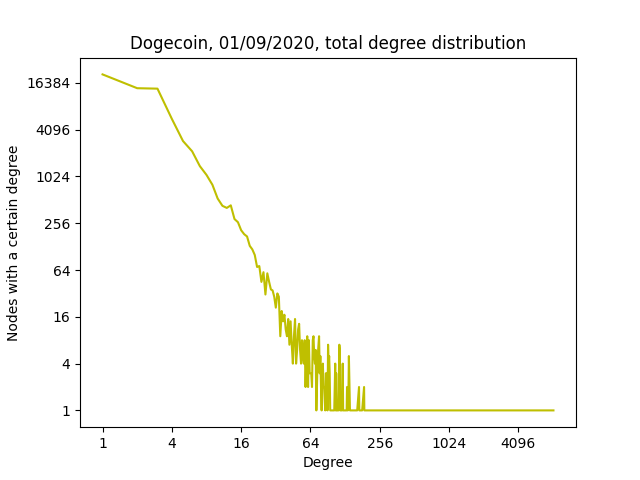}
    \includegraphics[width=0.5\textwidth]{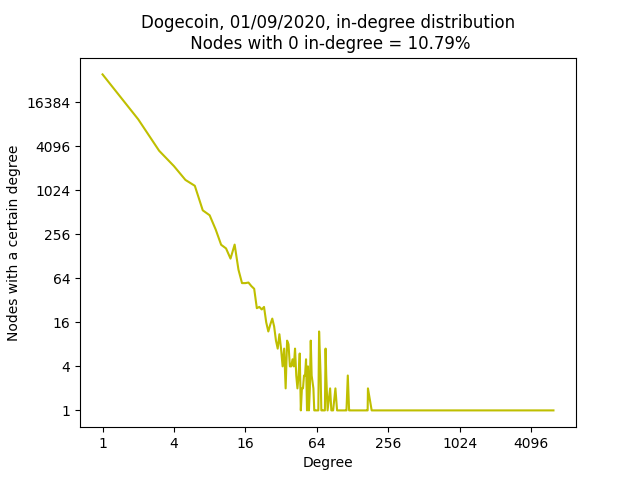}
    \includegraphics[width=0.5\textwidth]{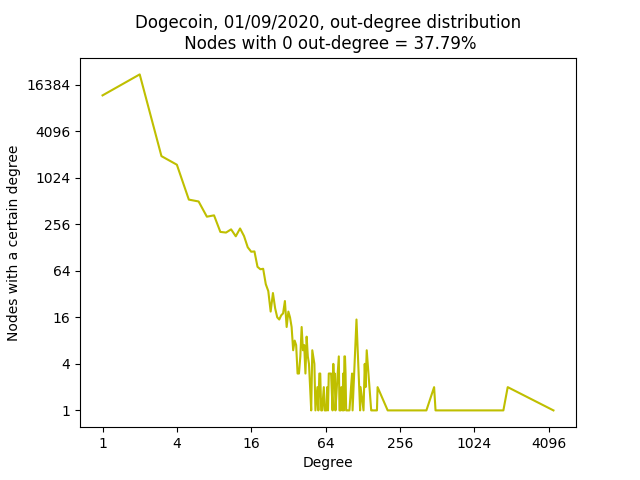}
  \caption{In, out and total degree distribution of the Dogecoin graph. Based on the transactions dated: 1st September 2020. The plot is performed in logarithmic scale on both axes.}
  \label{dogedegree}
\end{figure}

The degree distribution, shown in Figure~\ref{dogedegree}, has some peculiarities. First of all, the nodes with $0$ out-degree are almost four times more than the nodes with $0$ in-degree. Then, there are significantly more nodes with $2$ as out-degree than nodes with $1$ as out-degree. As expected, also in this case there are few hubs including $7$ nodes with a degree greater than $1000$ and the most connected node that is linked with the 12.3\% of the network and through which 21\% of the shortest paths pass (see Table~\ref{load_centrality_doge}).

\subsection{Outcome Analysis}
The results reported above show that the transaction graphs of all the analyzed cryptocurrencies have small-world properties, even though some of them have a more pronounced behaviour than others. 

\begin{table*}[h]
  \centering
  \caption{Comparison of the small world properties among the DLTs, one day and one month of transactions.} 
\begin{tabular}{|c|c|c|c|c|c|}
\hline
\textbf{Distributed Ledger} & \textbf{Nodes}  & \textbf{$|E|/|N|$}  &  \textbf{ASPL} & \textbf{ACC}  &  \textbf{$\sigma$}\\
\hline
\textit{ETH(normal, 1 day)} & {322\,467} & 1.17   & 0.13 & {14\,320} & {110\,154}\\ 
\textit{ETH (internal, 1 day)}  & {91\,340} & 1.04 & 0.3 & 192 & 630\\ 
\textit{ETH (internal, 1 month)}  & {1\,458\,543} & 1.07 & 0.25 & {30\,596} & {122\,003}\\ 
\textit{BTC (1 day)} & {886\,296} & 2.49 & 0.69 & {10\,923} & {15\,830}  \\
\textit{XRP (1 day)} & {9\,717} & 1.61 & 0.27 & 598 & {2\,215} \\
\textit{XRP (1 month)} & {94\,593} & 1.76 & 0.24 & {12\,512} & {52\,336} \\ 
\textit{DOGE (1 day)} & {67\,111} & 2.14 & 0.68 & 841 & {1\,242}  \\ 
\hline
\end{tabular}
    \begin{tablenotes}
        \item[1] 1) DLT  2) Number of Nodes 3) Edges-Nodes Ratio 4) ASPL Ratio 5) ACC Ratio 6) Sigma. 
    \end{tablenotes}
\label{results}
\end{table*}

Regarding one day analyses, Ethereum showed by far the lowest ASPL ratio ($0.13$), followed by Ripple ($0.27$), Dogecoin ($0.68$) Bitcoin ($0.69$). On the other hand, for which concerns the ACC ratio, Ethereum and Bitcoin had a value greater than $10\,000$, while Ripple and Dogecoin had a value between $500$ and $1000$. This is probably because the clusterization level generally tends to grow when the size of the graph increases. In fact Ripple, considering a full month of transactions instead of just one day, has an average clustering coefficient ratio over $10\,000$ as well.

Several factors must be taken into account when interpreting the results, such as:
\begin{itemize}
    \item The presence of change addresses in Dogecoin and Bitcoin.
    \item The role of the smart contracts in Ethereum (interactions among groups of users are performed through smart contracts, that thus become common network neighbors to all these users).
    \item The existence of exchange platforms that are connected to a lot of nodes.
    \item The presence of transactions with multiple inputs and multiple outputs in Bitcoin and Dogecoin, which also lead these cryptocurrencies to have a higher edges-nodes ratio with respect to Ethereum and Ripple.
    \item Finally, the common practice to adopt a new address or wallet for any novel transaction, in order to increase the users anonymity and unlinkability between transactions (or at least, reduce the ease to aggregate accounts and de-anonymize them).
\end{itemize}

Table~\ref{results} shows the mentioned ratio of the metrics, as well as the $\sigma$ value.

\begin{figure*}
  \centering
    \includegraphics[width=0.7\textwidth]{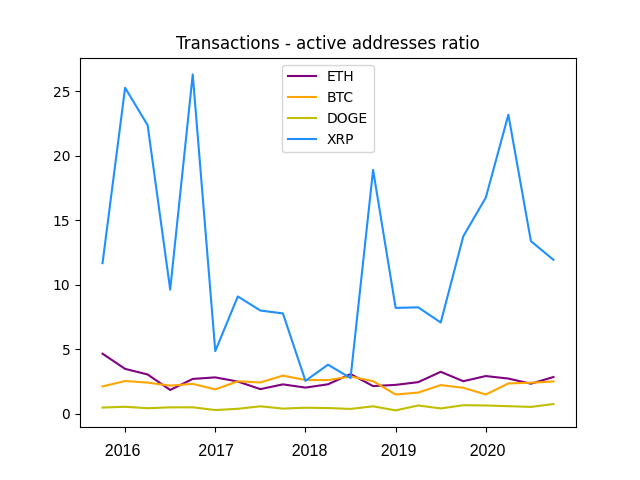}
  \caption{Transactions - active addresses ratio, data from the first day of each quarter since late 2015 (Jan 1st, Apr 1st, Jul 1st, Oct 1st).}
  \label{trans-addr-raio}
\end{figure*}

The different structuring of the systems brings, other than a different edges-nodes ratio, even to a bigger transactions-addresses ratio. Figure \ref{trans-addr-raio} shows that Ripple has considerably more transactions compared to the number of addresses involved with respect to other DLTs. In addition, Ripple payments-addresses ratio is very irregular, particularly because the number of involved accounts is subject to severe fluctuations. In Dogecoin, on the other hand, the number of active nodes is greater than the number of transactions.

All the analyzed graphs show the presence of hubs, while most of the nodes have very few connections. This is probably due to the presence of cryptocurrencies exchange platforms (e.g.~Binance), which interact with a lot of accounts that aim to convert fiat currencies into cryptocurrencies or vice versa. By observing the degree distribution results, we note that the plots in logarithmic scale look pseudo-linear. In fact, in all these charts the first part of the line representing the degree distribution well approximates a linear decade, thus suggesting a power law relation. Then, a tail is present with a minimal amount of nodes with much higher degrees (than others). This suggests that the considered networks might have a weak scale-free structure~\cite{Broido2019}.

\section{Conclusions}\label{sec:conc}
In this paper, we presented the Distributed Ledger Network Analyzer (DiLeNA), a new software tool designed for downloading the transactions recorded in certain Distributed Ledger Technologies (DLTs) and to compute a set of metrics on the resulting interaction graph. Based on the current version of DiLeNa, we studied four prominent DLTs: Bitcoin, Ethereum, Dogecoin and Ripple. Our analyses revealed that all the transactions graphs taken into account exhibit small world properties, although in certain cases such behaviour is more pronounced. Furthermore, from our tests it turns out that by stretching the period of time considered for the analyses (in our case from one day to one month) the small world behaviour becomes more marked in the observed graphs. This is because while the ACC of the random graphs tends to decrease when the size grows, the ACC of the transactions networks remains stable or even increases, causing the ACC ratio to be higher.

As reported before, the modular structure of DiLeNa permits to easily add the support for other DLTs, although it is not always possible to find a way to efficiently automatize the download of the transactions given a range of time. Another future extension of DiLeNa aims to further increase the computation parallelization throught GPUs. In this case, the main bottleneck is the lack of support for GPUs in the current version of the NetworkX library used for computing some of the metrics described in the paper. In other words, this will require us to switch to another library or to embed the metrics computation in DiLeNa using a more efficient programming language.

\bibliographystyle{spmpsci}
\bibliography{biblio}

\end{document}